\documentclass[twocolumn,twoside,slac_two]{revtex4}
\usepackage{graphicx}
\usepackage{fancyhdr}
\begin{document}

\title{The Interplanetary Network Supplement to the Fermi GBM Catalog - An AO-2 and AO-3 Guest Investigator Project}

%

\author{K. Hurley}
\affiliation{University of California, Space Sciences Laboratory, Berkeley, CA, U.S.A.}
\author{M. Briggs, V. Connaughton}
\affiliation{University of Alabama, Huntsville, AL, U.S.A.}
\author{C. Meegan}
\affiliation{USRA Huntsville, AL, U.S.A.}
\author{A. von Kienlin, A. Rau, X. Zhang}
\affiliation{Max-Planck-Institut f\"{u}r extraterrestrische Physik, Garching, Germany}
\author{S.  Golenetskii, R. Aptekar, E. Mazets, V. Pal'shin, D. Frederiks}
\affiliation{Ioffe Physico-Technical Institute of the Russian
Academy of Sciences, St. Petersburg, Russian Federation}
\author{S. Barthelmy, T. Cline(1), J. Cummings(2), N. Gehrels, H. A. Krimm(3)} 
\affiliation{NASA Goddard Space Flight Center, Code 661, Greenbelt, MD 20771, U.S.A.,
(1) Emeritus, (2) also Joint Center for Astrophysics, University of Maryland, Baltimore, MD 
 (3) also Universities Space Research Association, Columbia, MD}
\author{I. G. Mitrofanov, D. Golovin, M. L. Litvak, A. B. Sanin}
\affiliation{Space Research Institute, Moscow, Russian Federation}
\author{W. Boynton, C. Fellows, K. Harshman}
\affiliation{University of Arizona, Department of Planetary Sciences, Tucson, AZ, U.S.A.}
\author{R. Starr}
\affiliation{NASA Goddard Space Flight Center, Greenbelt, MD, U.S.A.}
\author{D. M. Smith}
\affiliation{Physics Department and Santa Cruz Institute for Particle Physics,
University of California, Santa Cruz, CA, U.S.A.}
\author{W. Hajdas}
\affiliation{Paul Scherrer Institute, 5232 Villigen PSI, Switzerland}
\author{K. Yamaoka}
\affiliation{Department of Physics and Mathematics, Aoyama Gakuin University, Sagamihara, Kanagawa, Japan}
\author{M. Ohno, Y. Fukazawa}
\affiliation{Department of Physics, Hiroshima University, Higashi-Hiroshima, Hiroshima, Japan}
\author{T. Takahashi}
\affiliation{Institute of Space and Astronautical Science (ISAS/JAXA), Sagamihara, Kanagawa, Japan}
\author{M., Tashiro, Y. Terada}
\affiliation{Department of Physics, Saitama University, Sakura-ku, Saitama-shi, Saitama, Japan}
\author{T. Murakami}
\affiliation{Department of Physics, Kanazawa University, Kadoma-cho, Kanazawa, Ishikawa, Japan}
\author{K. Makishima(1)}
\affiliation{Department of Physics, University of Tokyo, Bunkyo-ku, Tokyo, Japan,
 (1) also Makishima Cosmic Radiation Laboratory, The Institute of Physical and Chemical Research (RIKEN), Wako, Saitama, Japan}
\author{D. M. Palmer}
\affiliation{Los Alamos National Laboratory, Los Alamos, NM, U.S.A.}
\author{J. Goldsten}
\affiliation{Applied Physics Laboratory, Johns Hopkins University, Laurel, MD, U.S.A.}
\author{E. Del Monte, M.  Feroci}
\affiliation{IASF/INAF, Rome, Italy}
\author{M. Marisaldi}
\affiliation{INAF/IASF, Bologna, Italy}
\begin{abstract}
In the first two years of operation of the Fermi GBM, the 9-spacecraft Interplanetary Network (IPN) detected 158 GBM bursts with one or two distant spacecraft, and triangulated them to annuli or error boxes.  Combining the IPN and GBM localizations leads to error boxes which are up to 4 orders of magnitude smaller than those of the GBM alone.  These localizations comprise the IPN supplement to the GBM catalog, and they support a wide range of scientific investigations.
\end{abstract}

\maketitle

\thispagestyle{fancy}

\section{THE INTERPLANETARY NETWORK}

The IPN presently comprises AGILE, Fermi, RHESSI, Suzaku, and Swift, in low Earth orbit; 
INTEGRAL, in eccentric Earth orbit with apogee 0.5 light-seconds; Wind, up to 7 light-seconds from Earth; MESSENGER, in orbit around Mercury, up to $\approx$700 light-seconds from Earth, and Mars Odyssey, in orbit around Mars, up to $\approx$1200 light-seconds from Earth.  It operates as a full-time, all-sky monitor for transients down to a threshold of about $\rm 6\times10^{-7} erg\, cm^{-2}$, or $\rm 1\, photon \,cm^{-2}\, s^{-1}$, and detects about 325 cosmic gamma-ray bursts per year.  Due to the all-sky field of view and lesser sensitivity of the IPN, these bursts are generally not the same ones detected by more sensitive imaging instruments such as Swift BAT, INTEGRAL IBIS, SuperAGILE, and MAXI.  The IPN localization accuracy is in the several arcminute and above range.  The current burst detection rate of $\approx$325/year does not include magnetar bursts, to which the IPN is also sensitive.

We have now completed a preliminary analysis of the first two years of Fermi GBM data (July 14 2008 - July 14 2010).  In this period, the GBM reported approximately 500 GRBs.  Of them, 158, or about 32\%, could be triangulated using IPN data from one or more distant spacecraft, often in conjunction with Konus-Wind, to provide either a narrow error annulus or an error box.  A few examples are shown in figures 1, 2, and 3.  10 of the 158 were observed by the LAT; the IPN annuli have widths which are comparable to or less than the LAT error circle diameters (see figure 3).  30 of the 158 were independently localized by the Swift BAT or by Super-AGILE; these events are useful as end-to-end calibrations of the IPN.  

If the triangulation is coarse (several degrees) it can be used in conjunction with the GBM localization to produce a joint error box whose area is smaller than that of either one by itself.  When it is more accurate, it can also be used to refine the GBM systematic errors.  Since the IPN detects and localizes the stronger bursts, for which the GBM systematic uncertainties tend to dominate the statistical ones, IPN events are particularly useful for understanding these effects. This is analogous to the role which the IPN played in the BATSE era. 

IPN GRBs are being used to study polarization, to search for neutrinos ~\cite{achterberg},~\cite{stamatikos},~\cite{ahrens},~\cite{abbasi},
gravitational radiation  ~\cite{abbott1},~\cite{abbott2}, and VHE gamma-ray emission, to search for associations with supernovae ~\cite{pian},~\cite{hurley},~\cite{soderberg},~\cite{corsi},~\cite{walker}, and to determine whether high-B radio pulsars emit SGR-like bursts, among other projects.  Studies such as these do not require rapid localizations, or the identification of optical or X-ray counterparts, and constitute an alternative approach to the use of GRBs as astrophysical tools.  They benefit from using the large IPN database, which contains localizations of bursts which in general are more intense than those observed by imaging instruments, and therefore, on the average, closer (the redshifts IPN bursts range from 0.7 to 4.5, with an average of 1.6
and a median of 1.1).

\begin{figure*}[t]
\centering
\includegraphics[width=100mm]{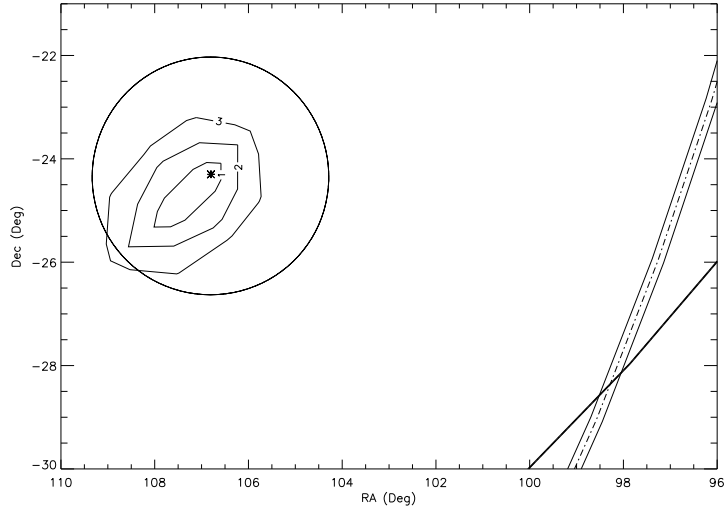}
\caption{Fermi GBM and IPN localizations of GRB 090228.  The contours are 1, 2, and 3 $\sigma$ confidence regions
derived from the GBM data.  The circle is an approximation to the 1 $\sigma$ contour, with a $2\,^{\circ}$ systematic
uncertainty added.  The asterisk indicates the most likely GBM position.  The narrow annulus is from Konus-Odyssey,
and the wide one is from Konus-MESSENGER.  Their intersection is the most likely IPN position.  The discrepancy
between the GBM and IPN positions may be due to large systematic uncertainties in the GBM localization.} \label{090228}
\end{figure*}

\begin{figure*}[t]
\centering
\includegraphics[width=100mm]{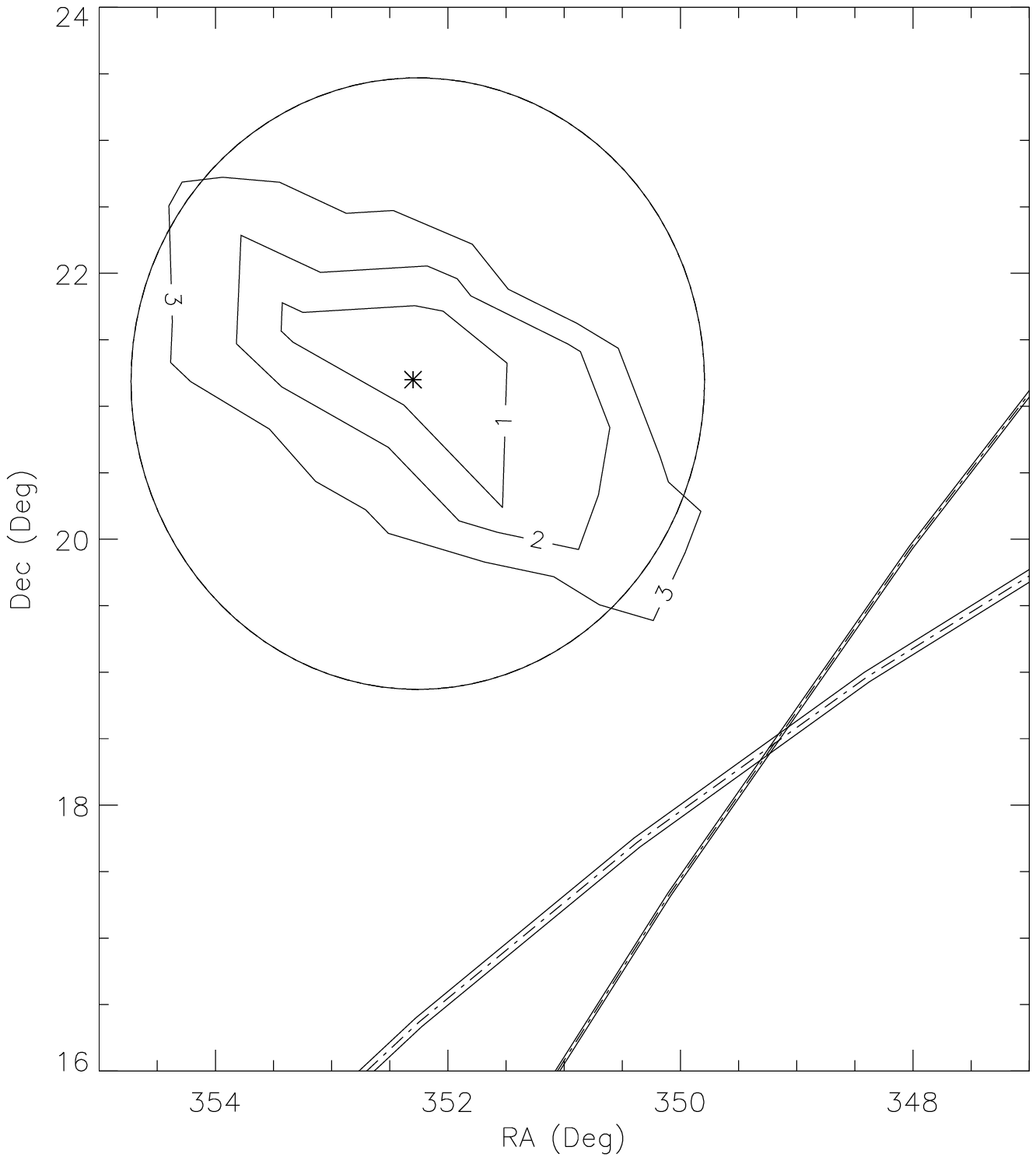}
\caption{Fermi GBM and IPN localizations of GRB 090131.  The contours are 1, 2, and 3 $\sigma$ confidence regions
derived from the GBM data.  The circle is an approximation to the 1 $\sigma$ contour, with a $2\,^{\circ}$ systematic
uncertainty added.  The asterisk indicates the most likely GBM position.  The narrow annulus is from Konus-Odyssey,
and the wide one is from Konus-MESSENGER.  Their intersection is the most likely IPN position.  The discrepancy
between the GBM and IPN positions may be due to large systematic uncertainties in the GBM localization.} \label{090131}
\end{figure*}

\begin{figure*}[t]
\centering
\includegraphics[width=100mm]{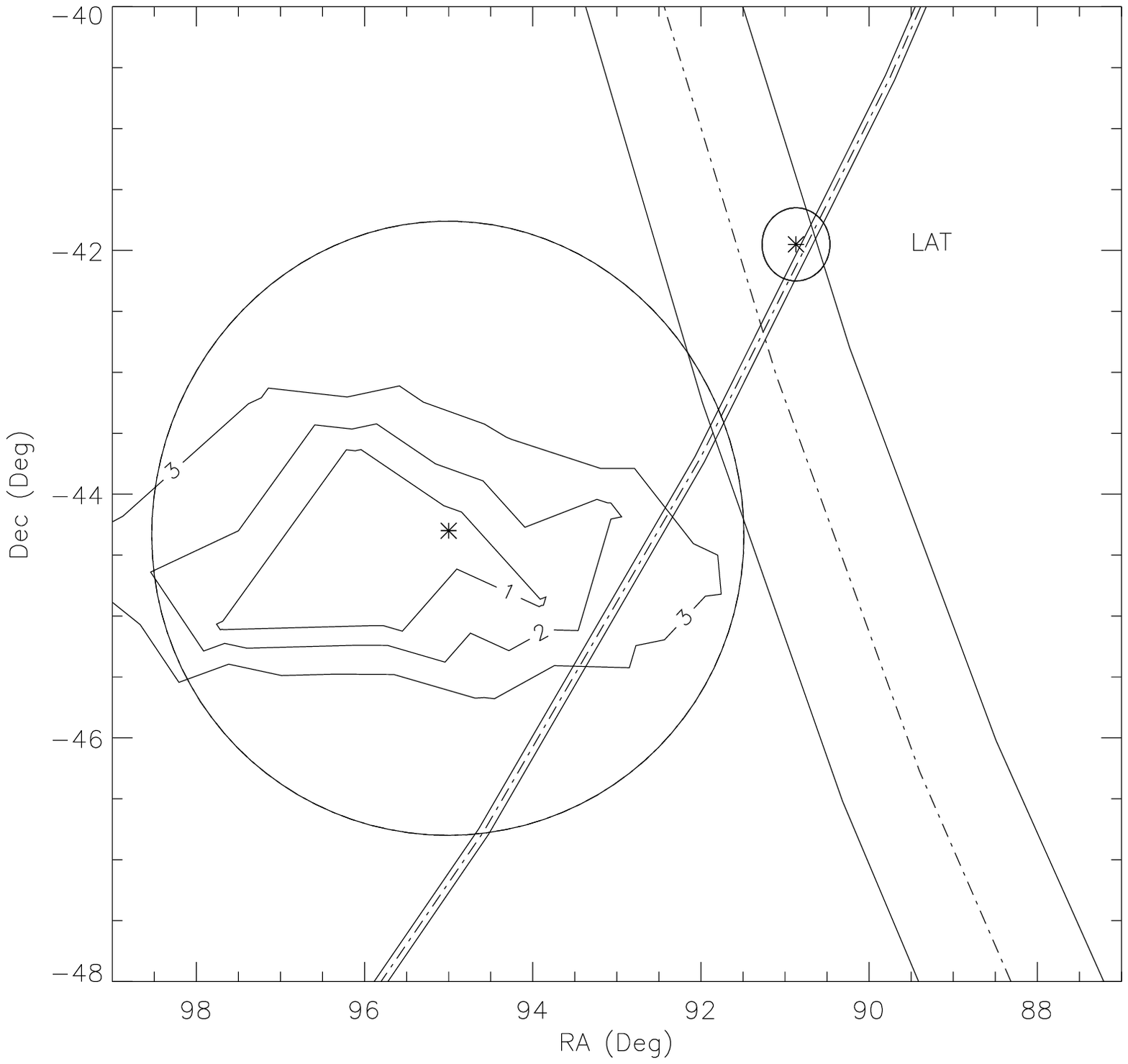}
\caption{Fermi GBM and IPN localizations of GRB 090328.  The contours are 1, 2, and 3 $\sigma$ confidence regions
derived from the GBM data.  The circle is an approximation to the 1 $\sigma$ contour, with a $2\,^{\circ}$ systematic
uncertainty added.  The asterisk indicates the most likely GBM position.  The narrow annulus is from Konus-MESSENGER,
and the wide one is from Konus-INTEGRAL.  Their intersection is the most likely IPN position.  The LAT error circle
is also shown.} \label{090328}
\end{figure*}

\begin{figure*}[t]
\centering
\includegraphics[width=100mm]{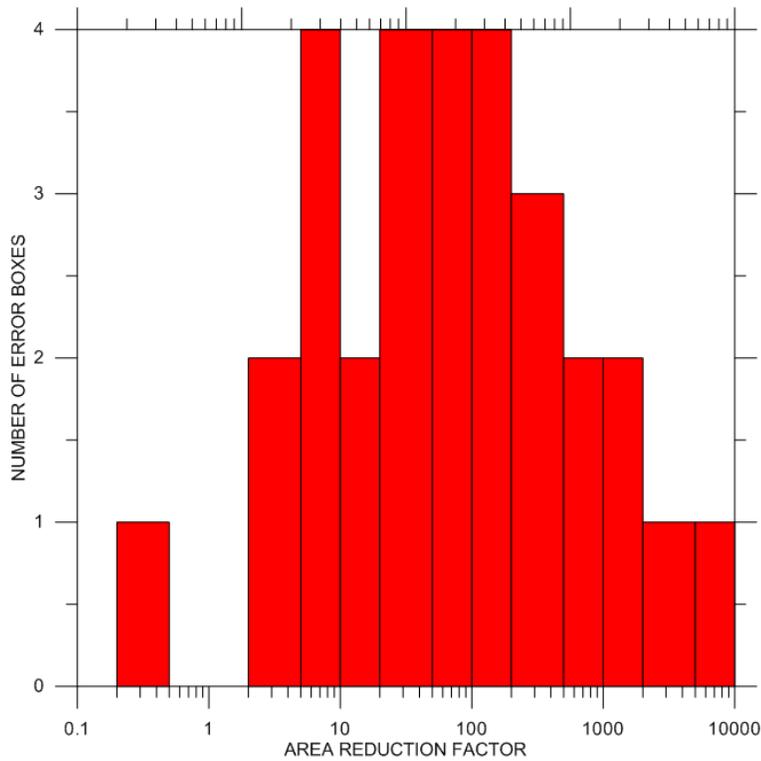}
\caption{The statistics of 30 IPN-enhanced GBM localizations.  The histogram gives the area reduction factor, defined as the ratio of the 1 $\sigma$ statistical-only GBM error circle to the 3 $\sigma$ IPN error box area.  Note that reductions of up to 4 orders of magnitude are possible.} \label{areareduction}
\end{figure*}

\begin{figure*}[t]
\centering
\includegraphics[width=100mm]{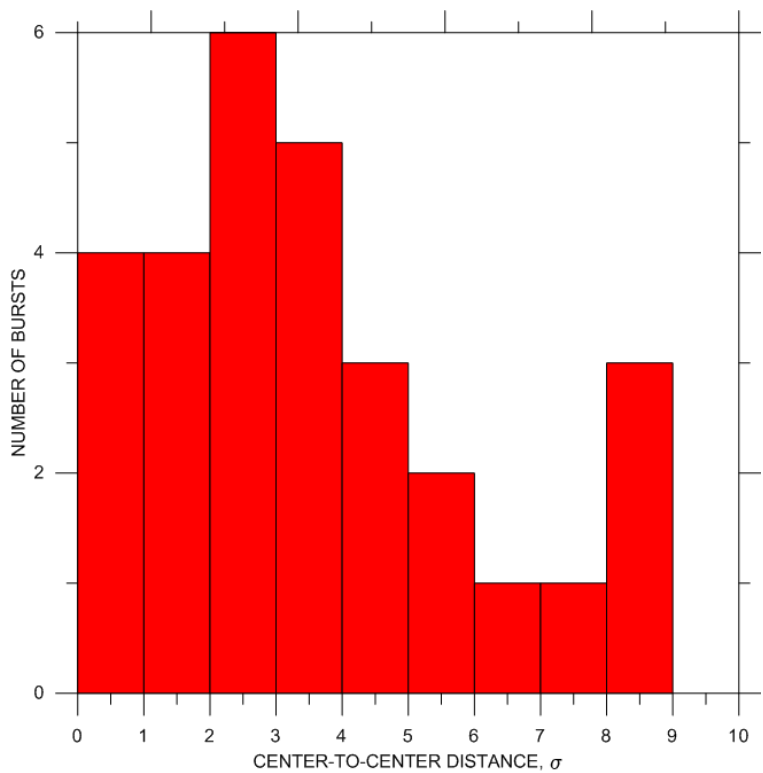}
\caption{The statistics of 30 IPN-enhanced GBM localizations.  The histogram gives the distance between the most likely GBM position and the center of the IPN error box, measured in units of the GBM 1 $\sigma$ (statistical only) error radius.  This illustrates the need for a systematic error component in some of the GBM localizations.} \label{centertocenter}
\end{figure*}

\begin{figure*}[t]
\centering
\includegraphics[width=100mm]{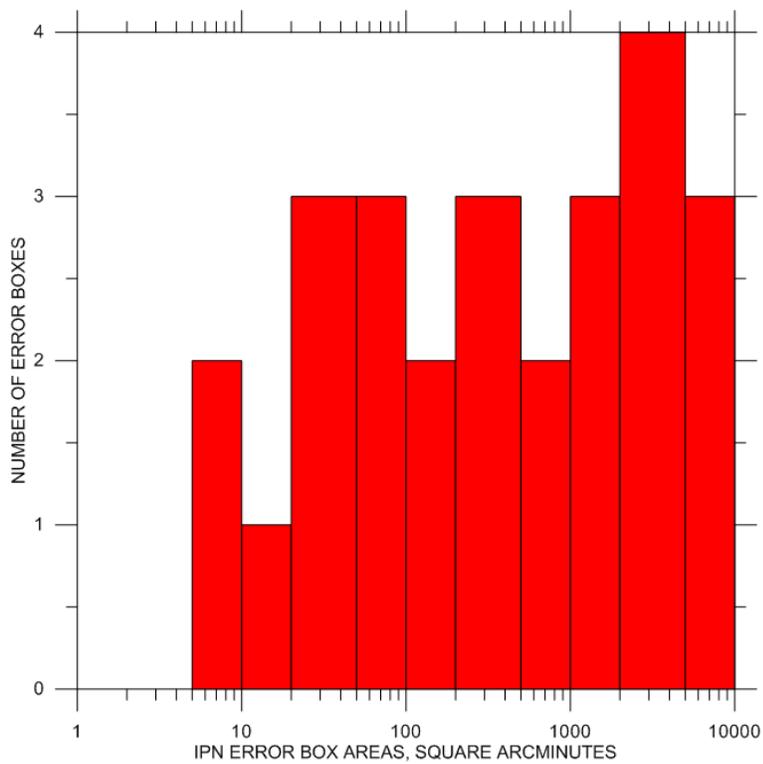}
\caption{The statistics of 30 IPN-enhanced GBM localizations.  The histogram gives the IPN 3 $\sigma$ error box areas.} \label{ipnareas}
\end{figure*}

\begin{figure*}[t]
\centering
\includegraphics[width=100mm]{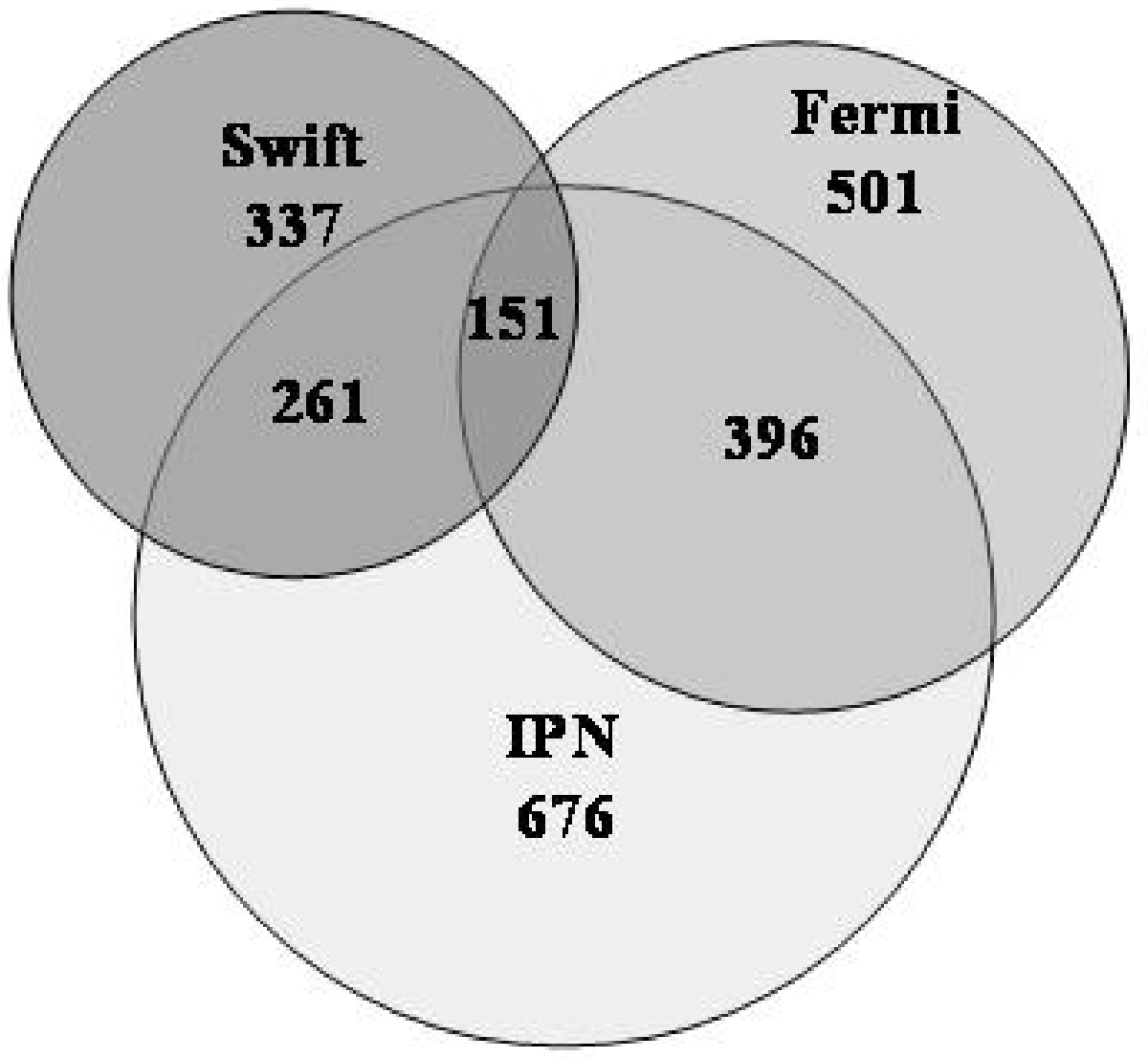}
\caption{Venn diagram (not to scale) showing the numbers of Swift, IPN, and Fermi bursts during the first two years of operation and their relation.  The Swift bursts are those both inside and outside of the BAT coded FoV.  IPN instruments observed 396 of the Fermi bursts, of which 158 involved one or more distant spacecraft (Odyssey or MESSENGER).} \label{ipnvenn}
\end{figure*}

\bigskip 
\begin{acknowledgments}
KH acknowledges IPN support under the following NASA grants: NNX10AU34G, NNX10AR12G, NNX10AI23G, NNX09AV61G, NNX09AU03G, NNX09AR28G, and NNX09AO97G.

\end{acknowledgments}

\bigskip 

\begin{thebibliography}{9}   



\bibitem{abbasi}
R. Abbasi et al., Ap. J. 710, 346, 2010

\bibitem{abbott1}
B. Abbott et al., PRL 101, 211102, 2008

\bibitem{abbott2}
B. Abbott et al., Ap. J. 681, 1419, 2008

\bibitem{achterberg}
A. Achterberg et al., Ap. J. 674, 357, 2008

\bibitem{ahrens}
J. Ahrens et al., Astropart. Phys. 20, 507, 2004

\bibitem{corsi}
A. Corsi et al., ApJ, 741, 76, 2011

\bibitem{hurley}
K. Hurley. \& E. Pian, E., AIP Conf. Proc. 937 (AIP: New York), p. 488, 2008

\bibitem{soderberg}
A. Soderberg et al., ApJ, submitted, 2011

\bibitem{stamatikos}
M. Stamatikos  et al., AIP Conf. Proc. 727 (AIP: New York), p. 146, 2004

\bibitem{pian}
E. Pian et al., in preparation, 2011

\bibitem{walker}
E. Walker et al., in preparation, 2011


\end{thebibliography}

\end{document}